\documentclass[graybox]{svmult}

\usepackage{mathptmx}       
\usepackage{helvet}         
\usepackage{courier}        
\usepackage{type1cm}        
\usepackage{makeidx}         
\usepackage{graphicx}        
\usepackage{multicol}        
\usepackage[bottom]{footmisc}

\usepackage{natbib}

\makeindex             

\begin{document}

\title*{Bulge growth through disk instabilities in high-redshift galaxies}
\author{Fr\'ed\'eric Bournaud$^1$}
\authorrunning{F.~Bournaud}

\institute{$^1$ Laboratoire AIM Paris-Saclay, CEA/IRFU/SAp, CNRS/INSU, Universit\'e Paris Diderot, F-91191 Gif-sur-Yvette Cedex, France. \email{frederic.bournaud@cea.fr}}
%
%
\maketitle

\abstract{
The role of disk instabilities, such as bars and spiral arms, and the associated resonances, in growing bulges in the inner regions of disk galaxies have long been studied in the low-redshift nearby Universe. There it has long been probed observationally, in particular through peanut-shaped bulges (Chapter~6.3). This secular growth of bulges in modern disk galaxies is driven by weak, non-axisymmetric instabilities: it mostly produces pseudo-bulges at slow rates and with long star-formation timescales. Disk instabilities at high redshift ($z$ $> $1) in moderate-mass to massive galaxies ($10^{10}$ to a few $10^{11}$\,M$_\odot$ of stars) are very different from those found in modern spiral galaxies. High-redshift disks are globally unstable and fragment into giant clumps containing $10^{8-9}$\,M$_\odot$ of gas and stars each, which results in highly irregular galaxy morphologies. The clumps and other features associated to the violent instability drive disk evolution and bulge growth through various mechanisms, on short timescales. The giant clumps can migrate inward and coalesce into the bulge in a few $10^8$~yr. The instability in the very turbulent media drives intense gas inflows toward the bulge and nuclear region. Thick disks and supermassive black holes can grow concurrently as a result of the violent instability. This chapter reviews the properties of high-redshift disk instabilities, the evolution of giant clumps and other features associated to the instability, and the resulting growth of bulges and associated sub-galactic components. 
}

\section{Introduction}
High-redshift star-forming galaxies mostly form stars steadily over long timescales, merger-driven starbursts being only a minority of galaxies (chapter~6.1; \citealt{lefloch}). At redshifts $z\,>\,1$, moderate-mass and massive star-forming galaxies ($10^{10}$ to a few $10^{11}$\,M$_\odot$ of stars) have rapid gas consumption timescales and stellar mass doubling timescales, of the order of a Gyr at $z\,=\,2$, depending mostly on redshift and weakly depending on mass, with rare deviations to the mean timescale (\citealt{schreiber}).

These star-forming galaxies have very irregular morphologies in the optical, especially compared to nearby spirals disks of similar mass, as unveiled by deep surveys over the last two decades. They also have very high gas fractions, about 50\% of their baryonic mass, as probed recently with interferometric studies. The high gas fractions and mass densities cause strong gravitational instabilities in the galactic disks, which results in disk fragmentation, and causes very irregular, clumpy morphologies. These irregular morphologies are often dominated by a few giant clumps of $10^{8-9}$\,M$_\odot$ of baryons, rotating along with the host galaxy.

This violent instability can drive the formation and growth of bulges, either by inward migration and central coalescence of the giant clumps and/or by gravitational torquing of gas and instability-driven inflows. This Chapter reviews the properties of the violent instabilities, and the pieces of evidence that the clumpy morphologies are caused by such violent disk instability rather than mergers or other processes. It then reviews the evolution of the giant clumps, their response to intense star formation and associated feedback processes, and the properties of bulges formed through this process. It eventually reviews recent results on other sub-galactic structures (such as thick disks and central black holes), which may grow concomitantly to bulges through high-redshift disk instabilities, and compares the role of this high-redshift violent instability to the contribution of low-redshift secular evolution through weak instabilities such bars and spiral arms.

\section{Clumpy galaxies and the violent disk instability at high redshift}

This section reviews the properties of star-forming galaxies at high redshift, the observed signatures of the underlying disk instabilities, and the main related theories. Throughout, we consider galaxies at redshift $z\, \approx \, 1-3$, with stellar masses of $10^{10}$ to a few $10^{11}$\,M$_\odot$, and that are ``normally'' star forming on the so-called Main Sequence \citep{daddi-ms,elbaz-ms,schreiber}, i.e. with specific star formation rates of the order of a Gyr$^{-1}$ at redshift $z\,=\,2$, as opposed to rare starbursts with faster star formation.

\subsection{Clumpy galaxies at redshift 1-3: global morphology}

The characterisation of the structure of star-forming galaxies at high redshift has steadily developed over the last two decades, driven mainly by deep surveys from the Hubble Space Telescope in the optical wavelengths \citep[e.g.][]{cowie96}, and later-on in the near-infrared, but also accompanied by modern techniques to identify and select star-forming galaxies in these deep surveys \citep[e.g.][]{daddi-bzk}. The highly irregular structure of high-redshift galaxies was first pointed out in the Hubble Deep Field \citep{abraham-hdf, vdb96, cowie96}. Star-forming galaxies appeared to have highly irregular morphologies, dominated by a few bright patches, with the striking example of the so-called ``chain galaxies'' where the patches are almost linearly aligned. While reminiscent of nearby dwarf irregulars, these morphologies where found in galaxies 10-100 times more massive -- a mass regime at which nearby galaxies are almost exclusively regular disk-dominated galaxies, most often barred spirals \citep{eskridge00,block02}, or spheroid-dominated early type galaxies. The lack of regular barred spirals, suspected in such deep optical imaging surveys \citep{vdb96,hdf-no-bars}, actually required deep-enough near-infrared surveys to be confirmed: otherwise the irregular structure could result from band-shifting effects, namely the fact that optical observations of $z\,>\,1$ objects probe the ultra-violet emission, strongly dominated by young star-forming regions, rather than the underlying mass distribution dominated by older stars. The first near-infrared surveys unveiled counterexamples of high-redshift galaxies with a more regular disk structure in the underlying older stellar populations \citep{sheth}, yet deeper and wider fields eventually confirmed the gradual disappearance of regular (barred) spiral disks at $z\,>\,0.7-1.0$ \citep{sheth08,comeron-bars,melvin,other-bars-candels}.

\begin{figure}[!ht]
\centering
\includegraphics[angle=270,width=9.5cm]{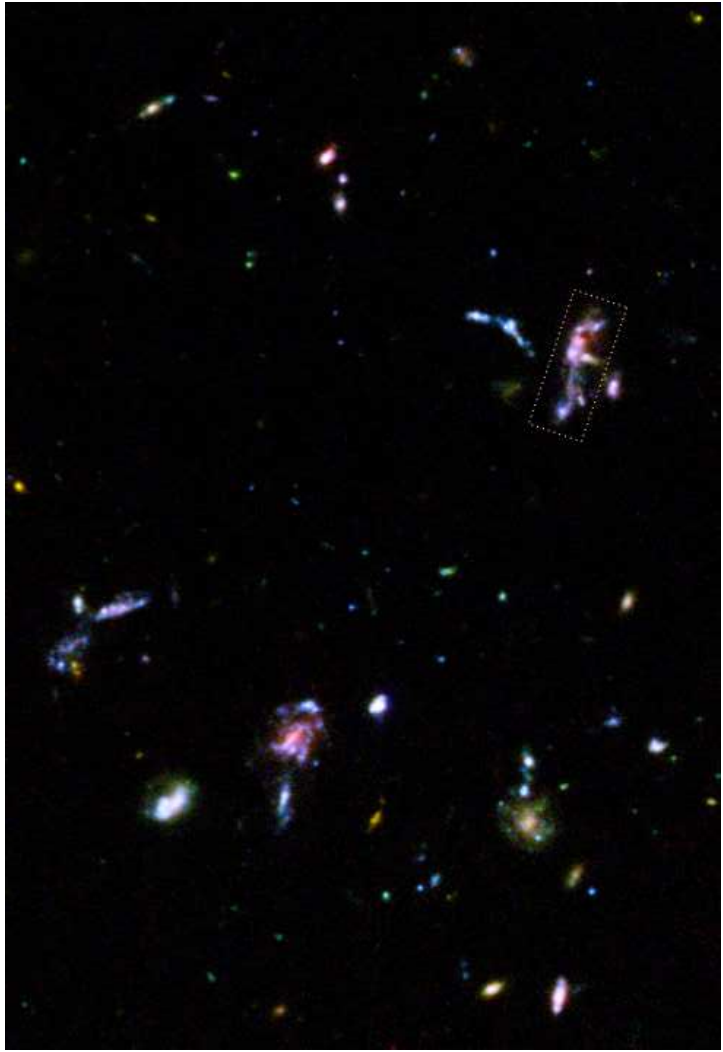}
\caption{Portion of the Hubble Ultra Deep Field (optical survey) showing more than 10 star-forming galaxies at $z=1-3$ with stellar masses of a few $10^{10}$\,M$_\odot$ (largest galaxies on the image). The galaxy in the dashed rectangle is a typical example of a Main Sequence galaxy about the mass of the Milky Way, located at redshift $z \simeq 1.6$. Its two apparent neighbours lie at very different redshifts. Like many galaxies in this mass and redshift range, it has a very irregular clumpy morphology, with a central reddish bulge and a few clumps of $10^{8-9}$\,M$_\odot$ each. Only the clump to the right of the centre is somewhat redder and contains old stellar populations, and only this particular clump might be a minor merger of an external galaxy. The others formed in-situ by gravitational instability in a gas-rich turbulent disk \citep{bournaud08} and follow a regular rotation pattern around the mass centre. The violent instability in the gas-rich medium can also trigger the asymmetry of the galaxy through an $m\,=\,1$ mode, without requiring an external tidal interaction.}
\label{fig_fb1}     
\end{figure}

Even before infrared data could resolve kiloparsec-scale structures at $z\,=\,2$, detailed spatially-resolved stellar population studies were used to reconstruct the stellar mass distribution of star-forming galaxies, in particular in the Hubble Ultra Deep Field (Figure~\ref{fig_fb1}). It was then found that the bright patches dominating the optical structure were not just random or transient associations of bright stars, but actually massive clumps with sizes in the 100-1000\,pc range, stellar masses of a few $10^8$ to, in extreme cases, a few $10^9$\,M$_\odot$, and typical stellar ages of a few $10^8$\,yr indicating relatively young ages (and in particular younger than those of the host galaxies) but suggesting lifetimes that are longer than their internal dynamical timescale (about 10-20\,Myr). As random associations would disrupt on such timescales, these data suggested that these bright regions were bound \citep{elmegreen05, elmegreen07}. These structures are generally dubbed ``giant clumps'', and their host galaxies ``clumpy galaxies'', although these are just the majority of star-forming galaxies at redshift 1--3 and the clumpiness is not a peculiar property of a rare type of galaxy. The morphology of clumpy galaxies suggested that these were disk galaxies, based in particular on the distribution of axis ratios from face-on to edge-on orientations \citep{E04,E06}, although the morphology of chain galaxies could have been consistent with filamentary alignments of separate small galaxies rather than edge-on clumpy disks \citep{taniguchi}.

\subsection{Kinematics and nature of clumpy galaxies}

More robust studies of the nature of clumpy star-forming galaxies were enabled by spatially-resolved spectroscopy of the ionised gas, probing the gas kinematics (velocity field, velocity dispersion) as well as chemical abundances and gradients in the interstellar medium. The pioneering study of \citet{genzel06} probed the disk-like nature of one typical star-forming galaxy at redshift two, with a disk-like velocity field, and no signature of an on-going or recent merger, in spite of an irregular and clumpy morphology. While in a single case a merger might show no observable kinematic signature if its velocity field looks like that of a disk, depending on the interaction orbit and on the observer's line-of-sight, much larger surveys have since then been assembled \citep{FS07, FS09, epinat12} and quantitative techniques have been used to interpret the velocity field structure and velocity dispersions of the observed systems \citep{shapiro08}. These confirmed that only a minority of clumpy galaxies display potential signatures from mergers, and that clumpier galaxies do not harbour more frequent signatures of recent or on-going mergers than smoother galaxies in the same mass range.

The velocity dispersions are high: typically about 50\,km\,s$^{-1}$ with large spatial variations for the H$\alpha$ gas (e.g., \citealt{genzel08, bournaud08}). Yet, they remain fully consistent with the large scale heights expected for these disks if chain galaxies, which are relatively thick, are the edge-on version \citep{E06}. There are also local irregularities in the velocity field, departing from pure rotation with non circular motions up to 20-50\,km\,s$^{-1}$, a few times larger than in nearby spirals, but again this is naturally expected from the presence of giant clumps: independently of their origin, massive clumps do stir the surrounding material by gravitational and hydrodynamic interactions, and they interact with each other: models of rotating galactic disks with massive clumps do predict such large non-circular motions on kilo parsec scales even when the disk is purely rotating before clump formation \citep{bournaud08}. 

Parametric classifications of galaxy morphology have long been based on low-redshift data and models tuned to represent the low-redshift Universe. Yet, such classifications now start to be tuned for for high-redshift galaxies, and optimised to avoid confusion between internal clumps and mergers. Such morphological classifications confirm the ``clumpy disk'' nature of the majority of Main Sequence star forming galaxies (Figure~2) and show close agreement with kinematic classifications \citep{cibinel}.

\begin{figure}[!ht]
\centering
\includegraphics[width=11.5cm]{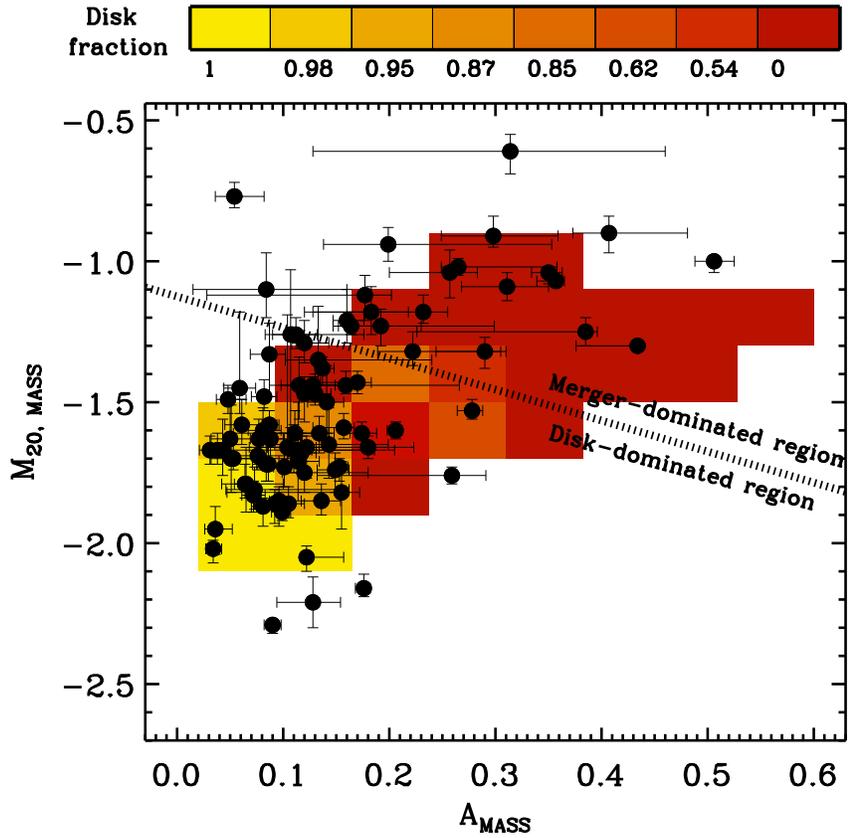}
\caption{Morphological classification of disks and mergers at $z \simeq 2$ (Cibinel et al. 2015). Combining the Asymmetry $A$ \citep{conselice} and the $M_{20}$ parameter \citep{lotz}, measured on stellar mass maps, is the most efficient way to distinguish clumpy irregular disks and genuine mergers. The probability of being a disk or a merger according to these parameters is coded using the background colours (note that the scale is logarithmic so disks strongly dominate all colour bins, except the last one). The black symbols show a mass-limited sample of star-forming galaxies at $z \sim 2$ in this $A-M_{20}$ plane (Cibinel et al. 2015). About two thirds of these objects are secure disks, and many others have a high probability of being a disk. The disk fraction is even larger when the sample is limited to Main Sequence galaxies, excluding the starbursts. Kinematic classifications by \citet{FS09} are in close agreement with this morphological classification, when applied to the same galaxies. Figure courtesy of Anna Cibinel.}
\label{fig_fb2}     
\end{figure}

Signatures of mergers might be more prominent in clumpy galaxies at intermediate redshift ($z\,=\,0.5-1$), according for instance to \citet{puech}. This is a different regime in which clumpy galaxies are more rare among star-forming galaxies, most of which have already started to establish a regular barred spiral structure (in the mass range considered here -- \citealt{sheth08, kraljic12}) and have much lower gas fractions and densities \citep{combes_fgas_interm}. Merger-induced clump formation may then become more prominent compared to higher redshifts. We also note that these clumps have lower masses so their dynamical impact may be weaker, their response to stellar feedback being likely different, and their role in disk and bulge evolution being potentially different as well. For these reasons the relatively rare clumpy galaxies below $z \simeq 1$ will not be considered hereafter.

\subsection{Observational insights on the nature of giant clumps: gas content and stellar populations}

The very nature of clumps and their formation process remain uncertain, in spite of the fact that their host galaxies are generally rotating disks, with a low frequency of mergers. Namely, are these structures formed in-situ, or do they originate from the outside in the form of small companion galaxies that have been accreted, or clumps of primordial gas that were accreted by the host galaxy before starting to form stars? 

The hypothesis of external star-free gas clouds might be ruled-out by the high average density of gas in the clumps (hundreds of atoms or molecules per cm$^3$, e.g. \citealt{E05}) making star formation very efficient, not counting the denser substructures likely to arise given the high observed turbulent velocity dispersions \citep{padoan99}.

This hypothesis of clumps coming from the outside as small companion galaxies joining a massive galactic disk through dynamical friction requires deep imaging to be examined through stellar population studies at the scale of individual clumps \citep{FS-imaging, wuyts12, E09}. The vast majority of clumps appear younger than expected for small external galaxies at the same redshift. Although their young stellar content of giant clumps may bias the age estimates by outshining the older stellar populations, the comparison with small galaxies at the same redshift shows that the clumps are significantly younger than small galaxies. Many clumps have estimated stellar ages of only about 100\,Myr with no underlying old stellar populations, while such populations would be detectable in small galaxies. 

If the clumps really form in-situ in their host galaxy, one should in theory sometimes capture the formation of the clumps during their first internal dynamical timescale ($\leq$\,20\,Myr) and hence some clumps should have extremely young ages (about 10\,Myr only). However, stellar population studies with broadband imaging and no spectroscopy cannot robustly distinguish such very young ages \citep{wuyts12}. Only recently such candidates could be identified with deep imaging and spectroscopy \citep{zanella}. At the opposite, considering galaxies with stellar masses of a few $10^{10}$\,M$_\odot$, a small companion of about $10^9$\,M$_\odot$ should be found within a projected distance of 10\,kpc for about one third of galaxies\footnote{This estimate is simply based on the mass function of galaxies and assuming a random geometrical distribution of satellites within the virial radius.}. If this satellite has not been fully disrupted by the galactic tides, its nucleus should be observed as a giant clump\footnote{With a kinematics that could become preferentially consistent with that of the host galaxy disk through gravity torques and dynamical friction within one galactic dynamical time, i.e. about 100\,Myr.}. Such ``ex-situ'' clumps, with older average stellar ages and an underlying old population, are indeed found in some cases. For instance the representative clumpy galaxy dissected in \citet{bournaud08} contains one clump which is much redder and older than the others, and also exhibits a larger deviation from the underlying disk velocity field, making external origin most likely for this one. Other candidates are found in \citet{FS-imaging} and appear in the statistics of \citet{wuyts12}. Nevertheless, such ex-situ clumps remain relatively rare: most observed clumps are actually different, with younger stellar ages \citep{E09, bournaud08, wuyts12}. The ability to identify such ex-situ clumps is actually reassuring that the non-detection of such old populations in the other clumps is robust, and probes their recent in-situ formation.

\medskip

Since most of the clumps formed recently inside their host galaxy, and given that they are gravitationally bound (based on the observed stellar masses and velocity dispersions, e.g., Elmegreen et al. 2005, Genzel et al. 2008), their formation likely involves a gravitational (Jeans) instability in a rotating disk, sometimes also called Toomre instability. For such instabilities leading to the formation of bound objects in a rotating disk to arise, the key requirement is a \citet{toomre} parameter $Q$  below unity\footnote{Although the Toomre $Q$ parameter is strictly meaningful only in an axisymmetric disk before strong perturbations arise. Note also that in a thick disk the instability limit is about 0.7 rather than $Q<1$ \citep[][and references therein]{behrendt}.}. Based on the observed rotation velocities and velocity dispersions (see previous sections) this typically requires gas density of the order of 100\,M$_\odot$\,pc$^{-2}$, an order of magnitude larger than in nearby disk galaxies, and implying interstellar gas masses comparable to the stellar masses, i.e. gas mass fractions of about 50\% of the baryonic mass. Such high gas densities had long been found only in the strongest starbursts galaxies (likely merger-induced) at any redshift \citep[e.g.,][]{tacconi08} but not in normally star-forming galaxies. The discovery that Main Sequence galaxies at redshift $z\,>\,1$ are actually very gas-rich with gas fractions of about 50\%, just counting the molecular content \citep{daddi08, daddi10, tacconi10, tacconi13}, has shed a new light on this issue. These high gas fractions are estimated from CO line observations, and hence subject to uncertainties on the conversion of CO luminosity to H$_2$ mass. Yet, new data probing the CO molecule spectral line energy distribution \citep{daddi15}, compared to detailed modelling of the CO excitation and emission in high-redshift galaxies \citep{bournaud15}, confirm high luminosity-to-mass conversion ratios and high gas mass fractions. Furthermore, the high gas fractions are also confirmed by independent estimates based on dust properties \citep{sargent, magnelli, genzel14}. 

Hence, the high inferred gas surface densities lead to Toomre parameters around unity for velocity dispersions of 30-50 km\,s$^{-1}$, and the Toomre $Q$ parameter would be even lower if the velocity dispersions were only a few km\,s$^{-1}$ as is the case in nearby spirals. Thus, axisymmetric gravitational instability should arise. All numerical experiments modelling disks with masses, sizes and gas fractions representative for the high-redshift star-forming galaxies consider here do show clump formation through gravitational instability \citep[see ][and the more detailed models discussed hereafter]{noguchi99, immeli, BEE07}. The gravitational stirring of the gas ensures that the turbulent velocity dispersions do not stay below the observed level of $\sim$50\,km\,s$^{-1}$: namely, the gas turbulent motions may also be powered by infall and stellar feedback, as we will review in the next sections, but at least the release of gravitational energy through the instability is sufficient to maintain high turbulent dispersions and self-regulate the disk at a Toomre parameter $Q \simeq 1$. 

The high velocity dispersions imply that the Jeans mass (or Toomre mass) is high, typically $10^{8}-10^{9}$\,M\,$_\odot$. This sets the high mass of the giant clumps forming through the associated instability. The properties of the instability in these disks was also modelled through analytic models by \citet{DSC09}, with results in close agreement to that of numerical models, and simpler analytic estimates. The giant clumps observed in high-redshift galaxies can thus be considered as the direct outcome and signature of the disk instability with a high characteristic mass, robustly expected from the basic observed properties (mass, size, rotation speed and gas fraction) of these high-redshift galaxies. This probably applies to most of the observed giant clumps, except the few ex-situ clump (minor merger) candidates with older stellar ages (see above). Note that the gravitational instability actually arises in a two-component disk with roughly half of its mass in gas, and half in stars: we refer the reader to \citet{jog-instab} and \citet{elmegreen-instab} for theoretical work on the two-component stability, and \citet{behrendt} for a detailed analysis of the disk stability in numerical simulations. The process is qualitatively unchanged compared to the $ Q \simeq 1$ self-regulated instability in a single-component disk described above; furthermore the gaseous and stellar velocity dispersions are probably about similar in these young galaxies (e.g., Bournaud et al. 2007) in which case the single-component stability analysis directly applies.

\subsection{The formation of gas-rich clumpy unstable galaxies in the cosmological context}

\begin{figure}[!ht]
\sidecaption
\includegraphics[angle=0,width=7.5cm]{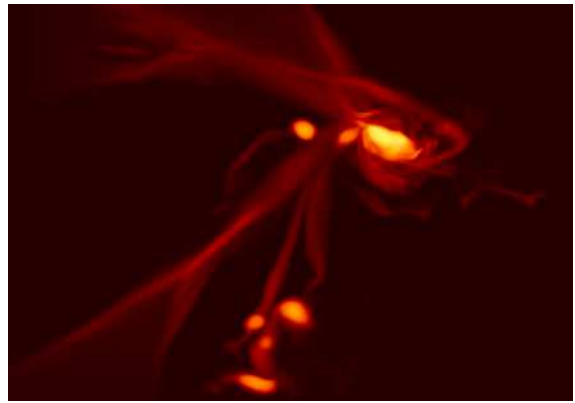}
\caption{Primordial galaxy fed by three cold streams of gas, in an idealised high-resolution simulation based on typical parameters measured in cosmological simulations \citep{gabor-sf}. The flows join the disk through a turbulent interface with extended circum-galactic reservoirs, before the gas dissipates its energy and feeds the cold star-forming disk, which keeps a high gas fraction, high velocity dispersions self-regulated at a Toomre parameter $Q\simeq1$, and a clumpy irregular morphology (image size: 100$\times$70\,kpc).}
\label{fig_fb3}     
\end{figure}

The irregular and clumpy structure of high-redshift star-forming galaxies is the outcome of their high gas fractions and densities. Theoretically, these high gas fractions are explained by the high rates of external mass infall, which are not compensated by excessively high star formation rates consuming the gas reservoirs, but rather preserved in a long-lasting steady state for Main Sequence galaxies. Recent cosmological models have highlighted the fact that high-redshift galaxies mostly accrete their baryons in the form of cold diffuse gas rather than hot reservoirs and companions galaxies \citep{dekel09, brooks}, which further helps the gas to rapidly join the cold star-forming disk. At the opposite, a too large contribution of galaxy mergers in the cosmological galaxy growth budget would form a massive stellar spheroid (bulge or stellar halo) too early and would stabilize $z\,\simeq \,2$ galaxies against giant clump formation, with a much lower turbulent speed and characteristic mass for any residual disk instability at $Q \simeq 1$ \citep{BE09}.
In detail, the cold accretion streams do not directly join the star forming disk (see Figure~\ref{fig_fb3}): they might be affected by the hot circum-galactic gas \citep{arepo} but more importantly need to dissipate the high kinetic energy from the infall, possibly by turbulent dissipation in circum-galactic regions \citep{elmegreen-burkert, gabor-sf} and/or forming extended rotating reservoirs that gradually feed the dense star-forming disk \citep{danovich}. Yet, detailed cosmological simulations show that part of the streams can directly feed the central few kiloparsecs of galactic disks in less than one dynamical time \citep{danovich}.

As a result of the high accretion rates, galaxies in cosmological simulations around redshift two have high gas fractions, although modern simulations still have difficulties to preserve sufficient gas reservoirs by avoiding excessive star formation at early epochs \citep[see][and references therein]{dekel-bathtub}. Nevertheless simulations with detailed stellar feedback models do produce some steadily star-forming galaxies with up to $\sim$ 30-40\% of gas at $z=2-3$\footnote{Although these {\em total} gas fractions seem to remain lower than the observed {\em molecular} gas fractions, especially if these are the most gas-rich galaxies in simulated samples.}. These simulations, further detailed in the next Section, display the expected gravitational instability for such gas-rich disks, and produce giant clumps and other dense features by gravitational instability in these disks \citep{agertz-clumps, ceverino10, ceverino12}. Only a limited fraction of the clumps are ``ex-situ'' clumps, resulting from the accretion of small nucleated companions or external gas clumps \citep{mandelker14}, fully consistent with the observations reviewed above.

\section{Mechanisms of bulge growth through high-redshift disk instabilities}

Knowing the properties of high-redshift star-forming galaxies from the previous Section, in particular the fact that they are subject to a violent clump instability rather than a weak axisymmetric instability as in nearby barred spirals, we now review the mechanisms through which bulge growth can be triggered and regulated by this instability mode.

\subsection{Clump migration and coalescence}

With a mass of a few $10^{8-9}$\,M$_\odot$, a giant clump in a high-redshift disk galaxy could behave just like a dwarf companion galaxy of similar mass, except that being dark matter free makes its mass distribution more spatially concentrated. It will in particular undergo dynamical friction on the underlying gaseous and stellar disk, and dark matter halo, and through this process will dissipate energy and angular momentum through increasing the velocity dispersion (i.e. the internal kinetic energy) of these massive components. In a rotating disk this will lead to inward migration of each giant clump until it reaches the galaxy centre -- like in a minor galaxy merger. 
In addition, as they lie in the disk plane, clumps also undergo gravity torques from other regions of the disk. A clump that forms in a purely rotating disk will break the symmetry of the mass distribution and induce a kinematic response in the form of a spiral arm or tidal arm denser than the average disk, in the disk plane \citep{bournaud-reviewmergers}. This over-dense region will then exchange angular momentum with the clump itself. If most of the mass lies at radii larger than the clump in the galactic disk, as is the case as soon as the disk is sufficiently extended radially, the strongest gravity torques will be from this arm onto the clump. Given that the outer disk has a slower angular velocity than the clump, this arm is trailing with respect to the rotation of the disk, so that the gravity torques exerted on the clump are negative, accelerating the inward migration already resulting from dynamical friction. This torquing process is most efficient in a gas-rich disk as the cold gas component makes the tidal arm response stronger than in a pure stellar disk.

As the process involves both dynamical friction and gravity torques, many numerical simulations have been used to study clump migration and estimate the migration timescale, starting with those of \citet{shlosman87} and those of \citet{noguchi99}, the latter being directly motivated by the first observations of chain galaxies and suggestions that the latter were internally fragmented disks by \citet{cowie96}. 

\begin{figure}[!ht]
\includegraphics[width=11.6cm]{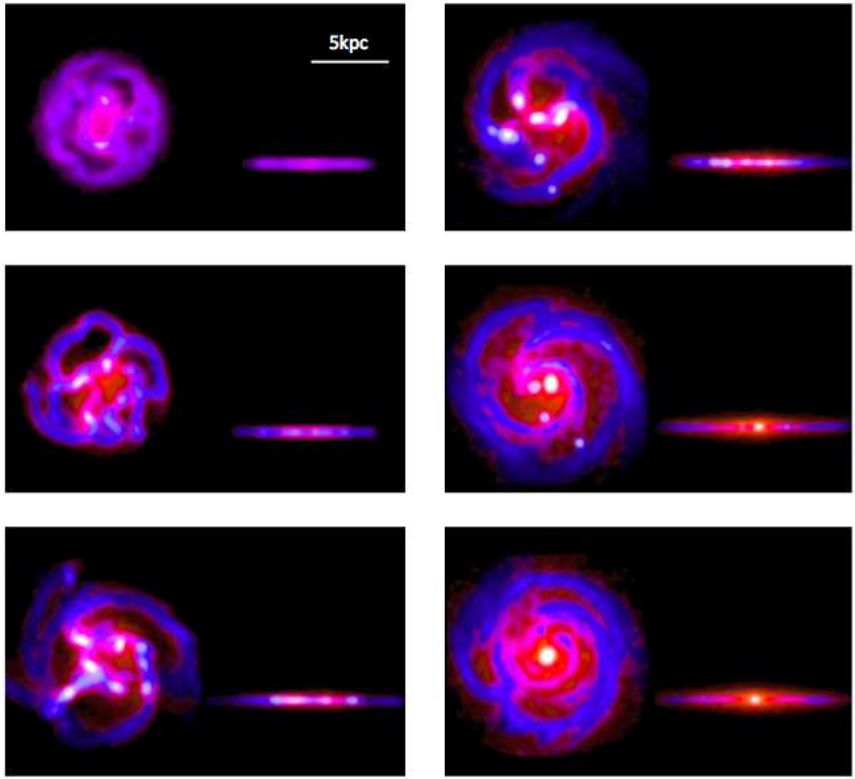}
\caption{Simulations of a gas-rich (50\% gas fraction) disk galaxy with initial parameters representative for star-forming galaxies at $z\,\simeq\,2$. The unstable gaseous disk forms a ring that quickly fragment into giant clumps. The clumps migrate inward and coalesce into a bulge, while the stellar disk is significantly thickened, and the disk radial profile, initially flat, is re-distributed into an exponential. The bulge formed here is a classical bulge with a S\'ersic index of 3.5--4.0. Blue codes gas-dominated regions and red codes star-dominated ones. Snapshots are separated by 100\,Myr. Simulation from \citet{BEE07}. }
\label{fig_fb4}     
\end{figure}

A more detailed treatment of the hydrodynamics and interstellar gas physics was introduced by \citet{immeli,immeli2}, and the simulations of Bournaud, Elmegreen \& Elmegreen (2007) were designed to finely correspond to the observed properties of star-forming galaxies in the Hubble Ultra Deep field at redshift $z\,=\,1-2$ with stellar masses of $10^{10}$ to $10^{11}$\,M\,$_\odot$ (see Figure~\ref{fig_fb4}). In the latter, the migration timescale of giant clumps to the galaxy centre is of the order of 300-500Myr, with a large scatter depending on the clump initial formation radius, and also of its interaction with other giant clumps and dense features in the disk. All of the experiments cited above are idealised models of isolated galaxies, lacking external replenishment of the disk by cosmological gas infall, accretion of smaller galaxies, and a few bigger mergers. As a result of star formation the gas fraction gradually decreases, and this can lead to over-estimating the clump migration timescale, as noted for instance by \citet{ceverino10}. However the clump migration process is so rapid (for clump masses above $10^8$\,M\,$_\odot$ at least) that the gas fraction decreases by less than a third of its initial value over this period, which is not larger than other uncertainties like the fact of not being able to observed any reservoir of atomic gas in these galaxies. Indeed cosmological simulations with external mass infall have reproduced the clump formation and migration process and find timescales that are consistent with the above, or only slightly shorter\footnote{Note that shorter migration timescales in cosmological simulations may also arise if the galaxies are too compact, or have too concentrated dark matter halos enhancing the dynamical friction process. Actually, the cosmological simulations of redshift two galaxies tend to have too low gas fraction because of some largely unexplained early consumption of the gas (e.g., Ceverino et al. 2012, Keres et al. 2012, with typical gas fractions at best around 30\% at $z\,=\,2$ even counting all the cold gas within a large radius).}.

\subsection{Possible evidence of clump migration}

An observational signature of inward clump migration, if they survive stellar feedback (see below), could be an age gradient with, on average, older clumps found at smaller radii. This is extensively quantified in the simulation sample of \citet{mandelker14}. In detail clump migration does not imply that young clumps cannot be found at small radii: external gas can feed high gas fraction in the inner disk \citep{danovich} and new clumps can form at small radii in gas-rich disks, except in the central 1-2\,kpc that are stabilised by strong shear and bulge mass \citep{BEE07,mandelker14}. The expected signature is rather an absence of aged clumps in the outermost disk, and clumps formed there have migrated inward in 100-200\,Myr (exceptions could still be found for moderate-mass clumps that can be scattered out to large radii by bigger clumps).

Observationally, statistical samples or resolved clumps remain limited, and their ages are hard to estimate. Not only the age of stars in a clump is not a direct tracer of its age (because clumps loose and re-accrete material, see next Section), but also, the stellar age estimators are strongly dependent on many parameters such as the assumed star formation histories, especially at high redshift \citep{maraston}. Nevertheless, an age gradient is tentatively observed by \citet{FS-imaging}, in quantitative agreement with clump migration and central coalescence within a timescale of at most 500\,Myr \citep[see also][]{guo-ages,guo14}.

\subsection{Stellar feedback, outflows, and the clump survival issue}

A key issue in the process of inward clump migration (and subsequent coalescence into a central bulge) is their response to stellar feedback. In nearby galaxies molecular clouds are estimated to have short lifetimes of the order of 10-20\,Myr, under the effects of supernovae explosions and other feedback processes \citep{gmc-lifetime, murray2}. While giant clumps are typically a thousand times more massive than the Milky Way biggest gas clouds, they are also ten times larger in each dimension, so their 3-D mass density is not necessarily much higher. As they form stars at high rates of a few M$_\odot$\,yr$^{-1}$ per clump \citep{E07,FS09,wuyts12} the released energy per unit gas mass is of the same order as that found in nearby star-forming clouds, or slightly higher. This raises the important question of clump survival against stellar feedback. In particular, having a released feedback energy per unit gas mass of the same order as in nearby molecular clouds does not mean that the giant clumps will be disrupted in a similar way or on a similar timescale: their gas also lies in a deeper gravitational potential well.

A first attempt to address this issue in \citet{EBE08bulge} concluded that if feedback was strong enough to disrupt the giant clumps within their migration timescale, it would also severely thicken the gas disk and heat the stellar disk well above the observed levels, almost disrupting any pre-existing rotation-dominated stellar disk. This was however based only on energetic supernovae feedback, while other stellar feedback mechanisms may be more likely to disrupt clumps in a realistic way, without disrupting the whole host galaxies. In particular, radiation pressure from young massive stars on the surrounding gas and dust may inject enough momentum into the clumps to disrupt the clumps \citep{murray}.

It has long remained difficult to address this issue in numerical simulations, mostly because stellar feedback can only be modelled through uncertain sub-grid models, even if modern hydrodynamic simulations of galaxies can reach sub-parsec spatial resolutions with mass resolution elements of the order of 100\,M$_\odot$ \citep{renaud13}. In fact, even the star formation rate which determines the powering rate of feedback relies on sub-grid models. Even if the star formation rate of entire galaxies or giant clumps is realistic compared to observations, changing the sub-grid model may significantly alter the spatial distribution of star formation, especially in resolution-limited simulations. A reassuring point is that state-of-the-art idealised simulations of galactic physics can now robustly model gaseous structures up to densities of $10^6$\,cm$^{-3}$ or more without being at their spatial resolution limit yet (i.e., the Jeans length at these densities can remain larger than a couple of resolution elements without necessarily requiring to bias the hydrodynamics by adding a temperature or pressure floor, Renaud et al. 2013). The fact that stars form with a quasi-universal efficiency in such dense gas \citep{krumholz, gao, garcia-b} implies that at least the first step of star formation, which is dense gas formation, is explicitly resolved in these simulations; the subsequent sub-grid modelling of star formation at fixed efficiency in high-density gas is consistent with observations down to scales much smaller than that of giant clumps -- this is now achieved in idealised simulations, but unfortunately remains out of reach of cosmological simulations so far. 

\begin{figure}[!ht]
\centering
\includegraphics[angle=0,width=11.6cm]{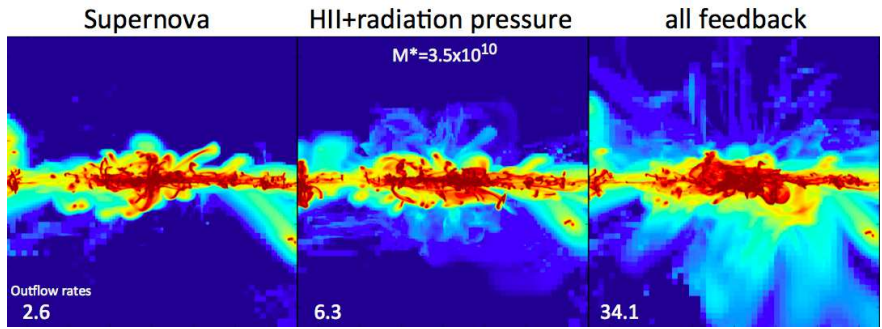}
\caption{Edge-on views of three simulations of the same gas-rich clumpy galaxy that has been evolved with different stellar feedback models during the last 80\,Myr (from left ro right: supernovae only, photo-ionization and radiation pressure only, and all mechanisms together, respectively). The gas density is shown and the outflow rates are indicated in the panels in M$_\odot$\,yr$^{-1}$ (measured 2\,kpc above/below the disk mid-plane). Outflows are launched by the giant clumps, and the models show the strongly non-linear coupling of feedback mechanisms: the total outflow rate in the simulations with all feedback processes together is well above the sum of the outflow rate in the independent cases. Similar non-linear coupling was noted by Hopkins et al. (2013). These simulations use the feedback models proposed by Renaud et al. (2013) and are similar to those presented in Bournaud et al. (2014), with 3\,pc spatial resolution.}
\label{fig_fb5}     
\end{figure}

The modelling of supernovae feedback is highly uncertain, in particular because it is often modelled through thermal dumps of the released energy heating the surrounding gas, while real supernovae remnants include a large fraction of the energy in non-thermal processes, which dissipate on slower timescales \citep{teyssier_sndelay}. Furthermore, models including the other kind of feedback processes such as stellar winds, photo-ionization, and most importantly radiation pressure were developed only recently \citep[][see Figure~\ref{fig_fb5}]{hopkins-feedback, renaud13} and include free parameters, an important one for radiation pressure being the number of scattering events that a photon can undergo in gas cloud before escaping from the cloud \citep{murray}. Another key parameter is the initial mass loading, namely whether the available energy or momentum is diluted into a large or a small mass (and volume) of gas. This parameter has long been unresolved in numerical simulations, and sometimes adjusted to generate ad hoc galactic outflows and study their fate \citep[e.g., ][]{oppenheimer, genel}. The highest resolution simulations of galaxies now become capable of resolving the typical distance over which photons from young stars redistribute their momentum into the ISM and start to estimate this loading factor from physical principles, even though explicit radiative transfer calculations robustly resolving these typical scale lengths are out of reach from galaxy-scale models and become feasible only in cloud-scale or clump-scale simulations.   

Some models of gas-rich galaxies with intense feedback have found that the giant clumps could be short lived, even with clump masses of the order of $10^9$\,M$_\odot$. This is the case for instance in the cosmological simulations from \citet{genel}, or the idealised models of \citet{hopkins-short-lived}. It is nevertheless remarkable that in these short-lived clumps models the clump lifetimes are very short, not larger than 50\,Myr, hence appearing inconsistent with the stellar ages estimated for real clumps, often reaching 100-200\,Myr and more (see above and Wuyts et al. 2012). In these models, the clump disruption is obtained in one or two generation of star formation and evolution, rather than through gradual, steady outflows on the longer term. Models with strong feedback and no long-lived clumps actually tend to lack giant clumps at all, strongly reducing and mass and/or number of clumps formed, at the point of being inconsistent with forming the majority of observed clumps by in-situ instability, as highlighted recently in the simulations of \citet{mayer}. However, in such models where in-situ clump formation is suppressed, a different (ex-situ) origin of clumps is not explained: in particular their stellar population ages can hardly be reconciled with minor mergers -- minor mergers can actually be identified as a sub-population of clumps that contain older stellar population \citep{bournaud08, E09} and these are only a small fraction of giant clumps. The suppression of in-situ giant clump formation obtained in the models of \citet{mayer} could in fact result of the low surface density of the disks in their initial conditions, which were inspired from cosmological simulations (which in turn may consume the disk gas too early), rather than from observed gas surface densities estimated through detailed analysis of the dust properties \citep{sargent,genzel14} and carbon monoxide spectral line energy distribution studies \citep{daddi15}. Hence, a common drawback of all theoretical models without long-lived clumps, is that either clump formation is suppressed, or the clump formation/disruption cycle is very short ($<$50\,Myr): in any case this appears inconsistent with the observations that commonly probe clump stellar ages of 100-200\,Myr and more. On the other hand, for long-lived clumps in a typical star-forming galaxy of stellar mass $10^{10-11}$\,M$_\odot$, the migration timescale from the clump birth site to the galaxy central kpc should be 300-500\,Myr, which appears slightly longer than the observed average stellar ages in giant clumps: this led Wuyts et al. (2012) to mention that clump disruption might be faster than clump inward migration. Yet, the clump stellar population ages provide only a lower limit to the clump real ages (see next paragraphs in this Section).

Actually, simulations with a thorough accounting of stellar feedback processes, including not just supernovae but also radiation pressure and other feedback mechanisms, do not necessarily predict short-lived clumps. In contrast with \citet{genel} and \citet{hopkins-short-lived}, models in \citet{perret14}, \citet{B14} or \citet{ceverino14a} include non-thermal and radiative feedback schemes and do find long-lived giant clumps -- at the same time they do correctly predict short lifetimes for gas clouds below $10^7$\,M$_\odot$ like in low-redshift galaxies. A different approach to feedback modelling by \citet{perez2013} also find long-lived clumps for any acceptable amount of stellar feedback, and even when strong outflows are launched by the giant clumps and their host galaxies, with outflow rates consistent with observations such as those of \citet{newman} and \citet{genzel11}. 

\medskip
 
Important constraints on the lifetime of giant clumps lifetimes and their ability to migrate inward toward bulges result from the fact that giant clumps are not quasi-closed-box entities, but steadily exchange mass with the surrounding medium in the host galaxy, through outflows and inflows of both gas and stars. Hence the ages of stars inside a given clump are not equal to the age of that clump since its actual formation: clumps have a sort of wave-like behaviour, although the pattern speed of the $m=0$ instability is almost equal to the disk rotation speed. Clumps may loose gas through stellar feedback, but more generally loose material through gravitational tides. At the clump half-mass radius, the gravitational force from the entire clump is only a few times larger than that from the entire galaxy. In other words, clump densities are only marginally higher than the limiting tidal density \citep{E05} and clumps gradually loose aged stars by dynamical evaporation toward the galactic potential well \citep{BEE07}.

Conversely, the clumps have a large cross section (of the order of 0.1--1.0\,kpc$^2$), and wander in a disk that contains substantial amounts of gas, even outside the giant clumps themselves\footnote{The presence of large amounts of gas between the giant clumps cannot be mapped spatially in CO surveys yet, but is predicted in the idealised and cosmological simulations of gas-rich unstable disks cited above, and confirmed by two observational arguments: (1) the emission from young stars in the ultraviolet contains a widespread component behind the giant clumps, tracing relatively dense gas \citep{E05} and (2) the CO spectral line energy distribution has two components, a high-excitation one attributable to dense clumps, and a low-excitation one corresponding to lower-density, large-scale background gas reservoirs \citep{daddi15, bournaud15}. }. Given this large cross-section of clumps, their relative velocity of 10--50\,km\,s$^{-1}$ with respect to surrounding gas, and a density of $\sim$10\,cm$^{-3}$ for the latter, accretion rates of 1--10\,M$_\odot$\,yr$^{-1}$ onto each giant clump are expected for pure ballistic capture. The gravitational potential well associated to the giant clumps may actually enhance the accretion. The first detailed estimates of this process were provided by \citet{KD13-accretion}. Detailed hydrodynamic simulations with an accurate AMR code \citep{teyssier02}, very high resolutions of 3--6\,pc, and detailed feedback models combining supernovae, photo-ionization and radiation pressure, were presented in \citet[][see also \citealt{perret14}]{B14}. These simulations confirmed that, independently on the details of stellar feedback and its ``strength'', clumps accrete a few solar masses per year of fresh gas. This gas accretion onto the clumps roughly compensates for both the gas consumption through star formation and the losses by gaseous outflows and dynamical evaporation of aged stars. This means that the clump actually evolves in a steady state, which can be described by a so-called ``bathtub model'' more commonly used for entire galaxies \citep{bouche12}: the gas infall rate is equal to the sum of the star formation rate and gas outflow rate, keeping the total mass constant, letting the system evolve in a steady state. The idea behind the steady state regulation is that any increase in the gas infall rate will be compensated for by the star formation rate, which has a non-linear response, and vice-versa for any decrease in the gas infall rate. The clump mass is then about stabilised, with some fluctuations, around its ``initial mass'', i.e. its baryonic mass after 1--2 times its own gravitational free-fall time (10--20\,Myr).

\medskip

An important prediction of the long-lived clump scenarios is that the average age of stars contained by a given giant clumps is younger than the actual clump age, measured since its formation by gravitational collapse of the gas-rich disk. The clumps experience a moderate starburst during their first 10-20\,Myr before feedback regulates star formation in a steady state regime \citep{zanella}. Then, they continue to form stars steadily at a higher rate than a closed-box system, due to \linebreak (re-)accretion of gas from the larger-scale galactic reservoirs. This keeps the average stellar age younger than the clump age. Furthermore, aged stars leave the clump gradually throughout the effects of dynamical heating and evaporation toward the galactic tidal field. As a result the stellar age is even younger than the clump age, and typical stellar ages of 100-200\,Myr are predicted for clumps of real age 300-500\,Myr, as the age of the stellar constant of giant clumps tends to saturate at 200-250\,Myr (Bournaud et al., 2014).

If clumps were disrupted by stellar feedback-driven outflows, the same processes of mass loss and accretion onto the clumps, which are driven by gravitational dynamics, would still be present. Hence the stellar ages would still set a firm minor limit to clump ages. Hence the observed stellar ages of giant clumps, typically of at least 100-200\,Myr \citep{wuyts12, guo-ages}, constitute evidence that clumps are not disrupted in a few tens of Myr as predicted by some feedback models -- actually, all the short-lived clump models reviewed above predict short lifetimes smaller than 50\,Myr. The observed stellar ages appear only consistent with models of clump survival and migration toward the galactic center, the ultimate limiting factor to the clump lifetime being their coalescence with other clumps or with the galactic bulge, explaining that Gyr-old giant clumps are not observed either.

\medskip

A typical star-forming galaxy at redshift $z=1-3$ can thus be expected to experience the migration and central coalescence of giant clumps of $10^{8-9}$\,M$_\odot$ of gas and stars\footnote{Note that the clumps remain gas-rich as they re-accrete gas and lose aged stars, which compensates for the gas depletion through star formation and gaseous outflows.}. Given the observed (and simulated) number of clump per galaxy, combined with their theoretical lifetimes and observed stellar ages, the central coalescence of a giant clump should typically occur at a rate of 10\,Gyr$^{-1}$ for a galaxy of stellar mass $10^{10-11}$\,M$_\odot$, i.e. one giant clump every $10^8$\,yr or so. If the unstable steady state lasts 2\,Gyr, this means that the baryonic mass reaching the bulge can be of order of $10^{10}$\,M$_\odot$ -- or even higher without strong stellar feedback, as in such a case the giant clumps have masses that increase through the accretion of surrounding gas without outflow regulation. 

\medskip

While the mass reaching the bulge can seem very high, we will see later that it does not necessarily entirely fuel a too massive bulge: in particular, a large fraction of the mass is still gaseous and this component can be expeled outward and/or form a central rotating disk rather than a bulge. In the next parts we first examine below the structural properties of the bulges formed by the central coalescence of giant clumps, and will review the bulge mass fraction issue subsequently.

\subsection{Instability-driven inflows}

\begin{figure}[!ht]
\centering
\includegraphics[width=11.5cm]{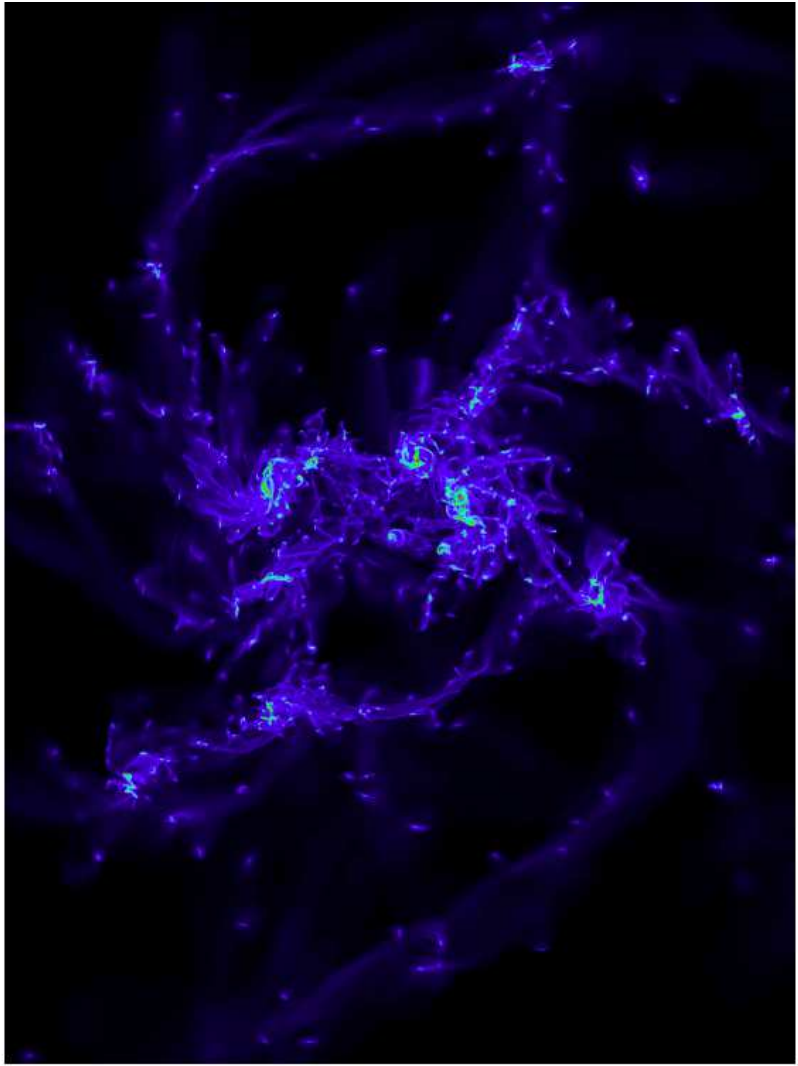}
\caption{Face-on view of the gas in a high-redshift galaxy simulation (image size: 8$\times$12\,kpc). The galactic rotation is counter-clockwise. Note the spiral armlets which are often on the leading side of giant clumps inside the clump radius, and on the trailing side in the outer disk. Gravitational torques from the clumps onto this inter-clump gas drive a continuous inflow of gas toward the galaxy center.}
\label{fig_fb6}     
\end{figure}

Another mechanism associated to giant clumps and disk instability, but different from giant clump migration and coalescence itself, can also grow the central mass concentration of high-redshift disk galaxies, and potentially the bulge mass. Giant clump and other dense features formed by gravitational instability exert gravity torques on the rest of the disk's gas, which transfers angular momentum. Clumps are located close to their own corotation radius (or slightly inside their corotation if dynamical friction has slowed down their rotation speed compared to the rest of the disk material). Material located at smaller radii in the disk thus rotates faster than the clumps, in terms of angular velocity. It then responds mostly as a leading tidal arm, found on the leading side of the clump compared to the galactic rotation. The presence of multiple clumps and other features can make the pattern hard to identify but a striking example with respect to one of the main clumps in a clumpy galaxy simulation is shown in Figure~\ref{fig_fb6}.

The material on the leading side of giant clumps undergoes negative gravity torques and loses angular momentum. The material in the outer disk gains angular momentum in exchange. The process is similar for any instability that breaks the disk symmetry \citep[e.g.,][]{combes-gerin, BCS05} but the gravity torques in the case of clump instabilities at high redshift are typically 10--20 times larger than for low-redshift secular instabilities (namely, spiral arms and bars): it can even transfer outward 100\% of the initial angular momentum in just one rotation period \citep{bournaud11}, compared to 5--10\% per rotation period for strong bars in low redshift spirals \citep{BCS05}. The corresponding mass inflow rate for a typical high-redshift star-forming galaxy is then of the order of 10\,M$_\odot$\,yr$^{-1}$ or more. 

The gravity torques between clumps and non-axisymmetric features are the main mechanism through which gravitational energy is pumped into the interstellar gas. As reviewed in the previous Section, these unstable disks evolve in a self-regulated regime at $Q \simeq 1$ with high velocity dispersions $\sigma \, \sim \, 50$\,km\,s$^{-1}$. Turbulent energy in the interstellar medium typically dissipates in a local crossing-time and more energy needs to be pumped in the turbulent cascade for steady-state systems \citep[e.g.,][]{maclow, bournaud11}. The specific energy loss rate is then $\sigma^2/(2\tau)$ where $\tau$ is about 10\,Myr, the energy being dissipated mainly through small-scale compression and shocks that heat the gas, which subsequently radiates the energy away. The radiative losses are balanced by the global inflow of gas down the galactic gravitational potential at an mass inflow rate $\dot M$, releasing an energy rate $\dot M V_c^2/2$ for a galactic circular velocity $V_c$. Typical outflow rates estimated for high-redshift star-forming galaxies are of the order of $\sim 10$\,M$_\odot$\,yr$^{-1}$ to compensate for the turbulent dissipation in a steady state: $\sigma^2/\tau \equiv \dot M V_c^2$ \citep{elmegreen-burkert, bournaud11, genel2}.

\medskip

Studies of the instability-driven inflow \citep[see for instance][]{krumholz-inflow,elmegreen-burkert,bournaud11} highlight the fact that the inflowing gas is not fully consumed by star formation. Actually, a large fraction of the initial inflow, several solar masses per year, typically flow onto the central kpc region or ``bulge region''. The gaseous inflow will not necessarily feed a classical bulge as the material is dissipative and discy, but feeds central star formation at a rate of a few solar masses per year. This is consistent with observations showing that this ``bulge region'' (about the central kpc) has younger stellar populations and more sustained star formation in the most unstable/clumpy galaxies than in smoother disks of similar mass and redshift \citep{E09,E11}. The instability-driven inflow thus increases the central concentration of gas {\it and} young stars, this stellar mass being potentially scattered into a pressure-supported spheroidal bulge during subsequent relaxation events (which can include: coalescence of other giant clumps, major interactions or minor mergers). Few simulations have studied in the outcome of the central mass concentration grown by this inflow, compared to direct bulge growth by central coalescence of clumps, mostly because the two processes would be hard to distinguish. Yet, it appears clearly that the stellar mass gathered in the central kiloparsec by the clump migration and instability-driven inflows ends-up in a bulge-like structure rather than just the innermost regions of a radially-concentrated rotating disk: it comes in excess of the disk exponential mass profile, and has high velocity dispersions dominating over the residual rotation \citep{EBE08bulge, inoue11, bournaud11}.

The absence of violent relaxation in the process of instability-driven gaseous inflows should produce only a so-called pseudo-bulge, namely a low S\'ersic index structure with substantial residual rotation. Yet, subsequent relaxation through clumps and mergers can make this mass contribute to a so-called classical bulge, i.e. a highly concentrated structure with virtually no angular momentum left. This global picture has not been studied in detail for disk-dominated galaxies with the mass of the Milky Way or up to $10^{11}$\,M$_\odot$, but it has been studied for more massive galaxies. As detailed in Section~5, these massive galaxies become compact spheroid-dominated throughout the violent disk instability process. The global instability driven inflow plays a major role in turning the initial disks into compact concentrated objects, while relaxation induced by giant clumps and some mergers turn them into ``classical'' spheroids (see also \citealt{zolotov}).

\subsection{Properties of bulges from high-redshift disk instability}

The properties of bulges resulting from the violent instability of high-redshift disk galaxies remain uncertain, as they largely depend on the lifetime and evolution of clumps against stellar feedback. If the clumps are short-lived, disrupted by feedback faster than their inward migration timescale, there is still a diffuse inflow of inter-clump gas driven by the instability (see above, Hopkins et al. 2012, and Bournaud et al. 2011), which will grow a low-concentration pseudo-bulge if not another relaxation process affects the central region. At the opposite, the models with long-lived clumps, more consistent with the observed ages and outflow rates, find that the instability can grow classical bulges, with a low rotational support and high S\'ersic indices after repeated clump coalescence \citep{EBE08bulge} and short star-formation timescales \citep{immeli2}. \citet{BEE07} have shown that for models scaled to the observed properties of galaxies in the Hubble Ultra Deep Field, the disk material is redistributed into an exponential profile during clump migration and classical bulge growth. 

Nevertheless, these first models included only supernova-like feedback schemes and lacked a more complete accounting of stellar feedback processes. The full series of stellar feedback processes regulates the mass and gas richness of clumps even if they remain long-lived. \citet{inoue2} proposed that the resulting bulge could rather be a pseudo-bulge, even if it is an old and metal-rich one. It nevertheless still seems possible to grow highly-concentrated classical spheroids, with detailed and thorough stellar feedback models, at least at high galactic masses \citep{ceverino14, zolotov}. The amount of relaxation was shown to be sufficient to form classical bulges with high S\'ersic indices at the center of exponential disks \citep{ceverino14}. Yet, strong subsequent evolution can occur and even basic parameters like the bulge:disk mass ratio can largely evolve between high-redshift unstable phases and present-day galaxies \citep{martig12}. \citet{bekki} highlighted some possible signatures of the past presence of giant clumps and contribution to present-day bulges. 

Thus, a consensus on the resulting bulge properties (mass and type) is far from being reached, recent efforts having mostly focused on understanding the nature of the giant clumps and their own evolution with respect to star formation and feedback. Note also that the central coalescence of clumps, while it may induce enough relaxation to produce classical bulges, can significantly erode a pre-existing dark matter cusp \citep{EBE08bulge, inoue-cusp}.

\subsection{Associated thick disk growth}
Another interesting mechanism associated to the instability of high-redshift disks is that pre-existing stars and stars that formed outside the giant clumps or have left the giant clumps, are rapidly scattered vertically by the local gravitational potential wells associated to the clumps themselves. The scale-height of the stellar disk thus rapidly increases into a very thick stellar system ($\geq 1-2$\,kpc). Even if the thin disk mass doubles by redshift zero and the old thick disk shrinks back by gravitational response (see \citealt{villalobos10}), a thick disk will remain with a typical scale height of 500-1000\,pc ; furthermore this instability-induced thick disk is decoupled from the younger thin stellar disk formed at lower redshift: this thick disk is a separate component in vertical profiles rather than just a low-density tail \citep{BEM09}.

An interesting property of the thick disks formed through this instability mechanism is that their growth is concomitant to bulge growth, possibly accounting for chemical similarities \citep{cc1,cc2}. Another noticeable property is the relatively constant thickness of the thick disk throughout the radial profile. While this fails to account for the outer flaring observed for thick disks, which is probably better explained by minor mergers and distant tidal interactions \citep{villalobos, dimatteo-thick}, it does successfully account for the presence of a thick disk in the innermost regions, around central disks and bulges, as observed \citep{dalcanton-bernstein}. This latter property could not be explained by minor mergers and tidal interactions, which stir and thicken preferentially the low-density outer regions of the stellar disk \citep{villalobos, BEM09, martig12, dimatteo-thick}. At the same time, clump instability cannot account for all observe thick disk properties and a contribution of other processes such as minor interactions and mergers is likely required, too \citep{inoue-thick}.

Therefore, while interactions and mergers appear needed to explain the outer structure of thick disks, clumpy disk instabilities at high redshift appear required to explain their inner one. An interesting property of thick disks is that the fraction of the stellar mass that they gather is larger in later-type galaxies with small bulge fractions \citep{joachim}. This relation might be explained by clump-driven bulge growth and disk thickening, if the early-formed stellar mass is distributed between the bulge and the thick disk, with preference for the bulge at high total mass and preference for the thick disk at lower total mass. The relation between bulge properties and thick disk properties, potentially resulting from the role of high-redshift disk instabilities in bulge growth, was further outlined by \citet{comeron} who argue for concurrent growth of the thick disk and central bulgy mass concentrations in the past history of today's spiral galaxies.

\subsection{Is bulge formation too efficient? Stellar feedback and bulge growth regulation}

A key question related to bulge formation or growth by disk instabilities is whether this mechanism would over-predict bulge formation. Standard $\Lambda$-CDM galaxy formation models already tend to over-produce bulges and spheroids, at the expense of high angular momentum disks, especially in the baryonic mass range of $10^{10}-10^{11}$\,M$_\odot$. At best, some baryonic physics models may result in an acceptable distribution of stars among bulges and disks \citep{agertz-cosmo, eris-sim} but generally over-produce the stellar mass \citep{guo-masses}, while models with more realistic stellar masses remain too dominated by bulges and low angular momentum components \citep{scannapieco, guo-masses}. While these results are often in tension with observations, or can at best be marginally reconciled, they are generally supported by semi-analytic models that do not include disk instabilities, or only low-redshift secular instabilities (bars) that grow bulges much more slowly \citep[e.g.,][]{sommerville-sam-instab}, or cosmological hydrodynamic simulations that generally do not resolve giant clumps and other features associated to disk instabilities, as this requires high resolution that is too costly to maintain down to $z=0$ \citep{ceverino10} -- some hydrodynamic models also employ thermal models and/or feedback schemes that suppress strong disk instabilities, as these were not observed until recently \citep{sommerville-clump-suppr}. Note also that cosmological simulations tend to overproduce stars at early epochs ($z$ $>$ 3) and preserve too low gas fractions down to redshifts $z=1-3$ (\citealt{dekel-bathtub} and references therein) which can also damp the disk instability process in these simulations.

The new mechanism of bulge formation by disk instability thus comes on top of a cosmological model which, depending on the assumed (and still uncertain) baryonic physics, already produces enough stellar mass in bulges and low angular momentum components -- if not already too much. This could be an indirect argument against clump survival and coalescence into bulges, although the observed clump ages are consistent with long lifetimes and migration to bulges. Note nevertheless that the instability driven inflow (see above) is independent of clump survival so its contribution to the growth of central compact components could not be suppressed in the case of short-lived clumps. The question of whether the proposed high-redshift disk instability mechanisms overproduce bulges is thus naturally raised. The typical numbers for a galaxy of stellar mass about $5\times 10^{10}$\,M$_\odot$ are, say, five clumps of $5\times 10^{8}$\,M$_\odot$ each, migrating to the bulge in 400\,Myr, with unstable steady state maintained for 2\,Gyr. This means that 25\% of the total stellar mass coalesces into the bulge through this process, leading to an excessive bulge:disk mass ratio of 1:3. Furthermore the diffuse instability-driven inflow may double this estimate, which was based based on the migration and coalescence of giant clumps only. The resulting bulge fraction could thus be above the acceptable levels for such moderate mass galaxies, and even if the thin disk doubles its mass without further growth of the bulge between $z=1-2$ and $z=0$, the resulting bulge:disk ratio at $z=0$, of at least 1:6, still seems in tension with observations -- especially if other processes, such as minor mergers and some bigger interactions that are unavoidable at some level, also grow the central bulge. These simple estimates highlight the potential issue. 

\medskip

Early models of clump formation and migration clearly over-produced bulge masses, with final bulge:disk mass ratios about 1:1 after the violent instability period \citep[e.g.,][]{noguchi99, BEE07}. A strong limitation of these models was the lack of stellar feedback other than weak supernovae feedback. While the global star formation rate of the galaxies were somehow regulated to realistic values on the Main Sequence, the clumps did not produce gaseous outflows at realistic rates close to their own star formation rates. As a result, not only they keep all their initial mass, but they accrete surrounding material and their cross-section for capture of new material only increases. While the initial clump mass in these models is in agreement with observations (with typical masses of a few $10^8$\,M$_\odot$ for the main few clumps, and rarely more extreme cases) during the first 50-100\,Myr, their masses eventually become excessive with frequent clumps above $10^9$\,M$_\odot$, and the resulting bulge masses after clump coalescence are too high as well.
 
\medskip

The realisation that clumps actually have their mass content regulated by feedback, with outflows compensating for the sustained gas accretion, helps solve the problem in two ways. First, the clump masses are regulated to a value fluctuating around their initial mass throughout their migration in the disk, reducing the mass available for bulge coalescence by a factor of a few. Second, the fact that clumps gradually lose their aged stars and re-accrete gas means that they remain gas-rich and even gas-dominated throughout their lifetime (Bournaud et al., 2014 -- without these processes they would accumulate large amounts of stars and become star-dominated before reaching the bulge region). More than half of the material reaching the bulge region is thus gaseous, and will form a rotating disk component, which will gradually turn into a stellar component later on and could be scattered into a bulge component subsequently, but also remains sensitive to feedback processes and may be ejected from the central regions through stellar or AGN-driven outflows. Detailed simulations quantifying the bulge growth for long-lived clumps having their mass regulated by supernovae, photo-ionization and radiation pressure feedbacks, show that the growth of bulges by clump migration and instability-driven inflows remains at fully acceptable levels, at least after 0.5-1.0\,Gyr of evolution, for galaxies of baryonic mass of $10^{10-11}$\,M$_\odot$ (see Bournaud et al. 2014 for quantitative results).

Whether a long-lasting unstable steady state will eventually over-produce bulges, or whether feedback processes can prevent too much gas to accumulate in the central kpc and turn into stars that will subsequently be scattered into a spheroidal bulge, remains an open question. Stronger regulation than just clump outflows appears needed if the clumps are actually long-lived and can migrate. A first solution to this problem was proposed by Perret et al. (2013). As clumps remain gas-rich along their evolution, their central coalescence conveys large amounts of gas inwards (typically a few $10^8$\,M$_\odot$ of gas over $\sim$0.1\,kpc$^2$) which provokes a local starburst in the central kpc. The resulting feedback expulses large amounts of gas -- from the coalescing clumps, as well as gas brought inward by the global instability-driven inflows. Gas expulsion rates from the central kpc peaking at a few tens of solar masses per year are reported by Perret et al. In addition, this process will affect any pre-existing stellar bulge. The gas mass indeed represents roughly half of the mass in the central kpc (dark matter providing only a minor contribution at such scales) and this major component rapidly fluctuates by clump inflows and feedback-driven gas expulsion. The orbits of stars in the bulge are affected by the rapid fluctuations of the gravitational potential -- a process already pointed out by \citet{bois} regarding stellar scattering by young star clusters and gas clouds in mergers, but involving much higher masses at high redshift. 

\begin{figure}[!ht]
\centering
\includegraphics[width=7cm]{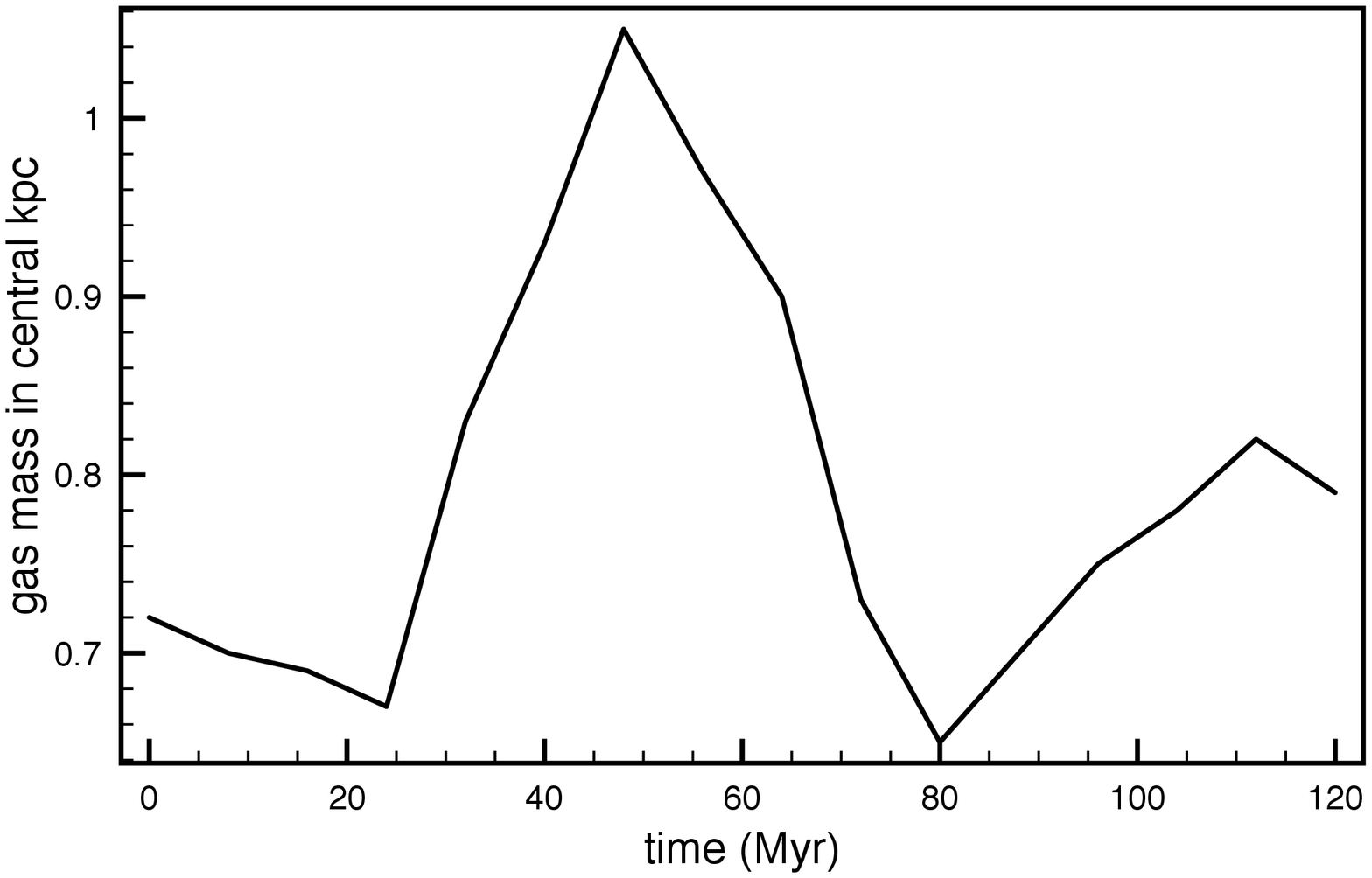}
\includegraphics[width=7cm]{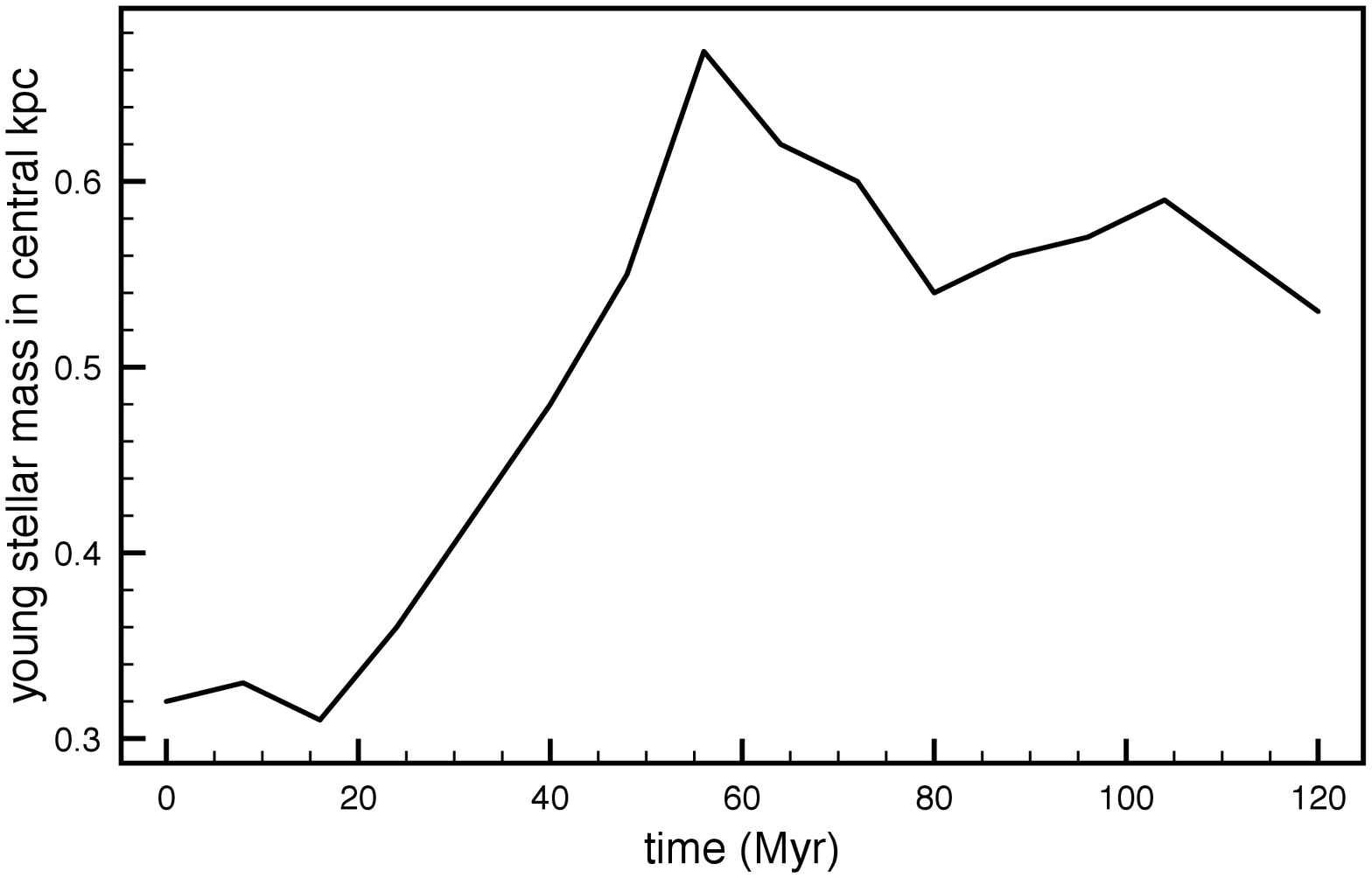}
\includegraphics[width=7cm]{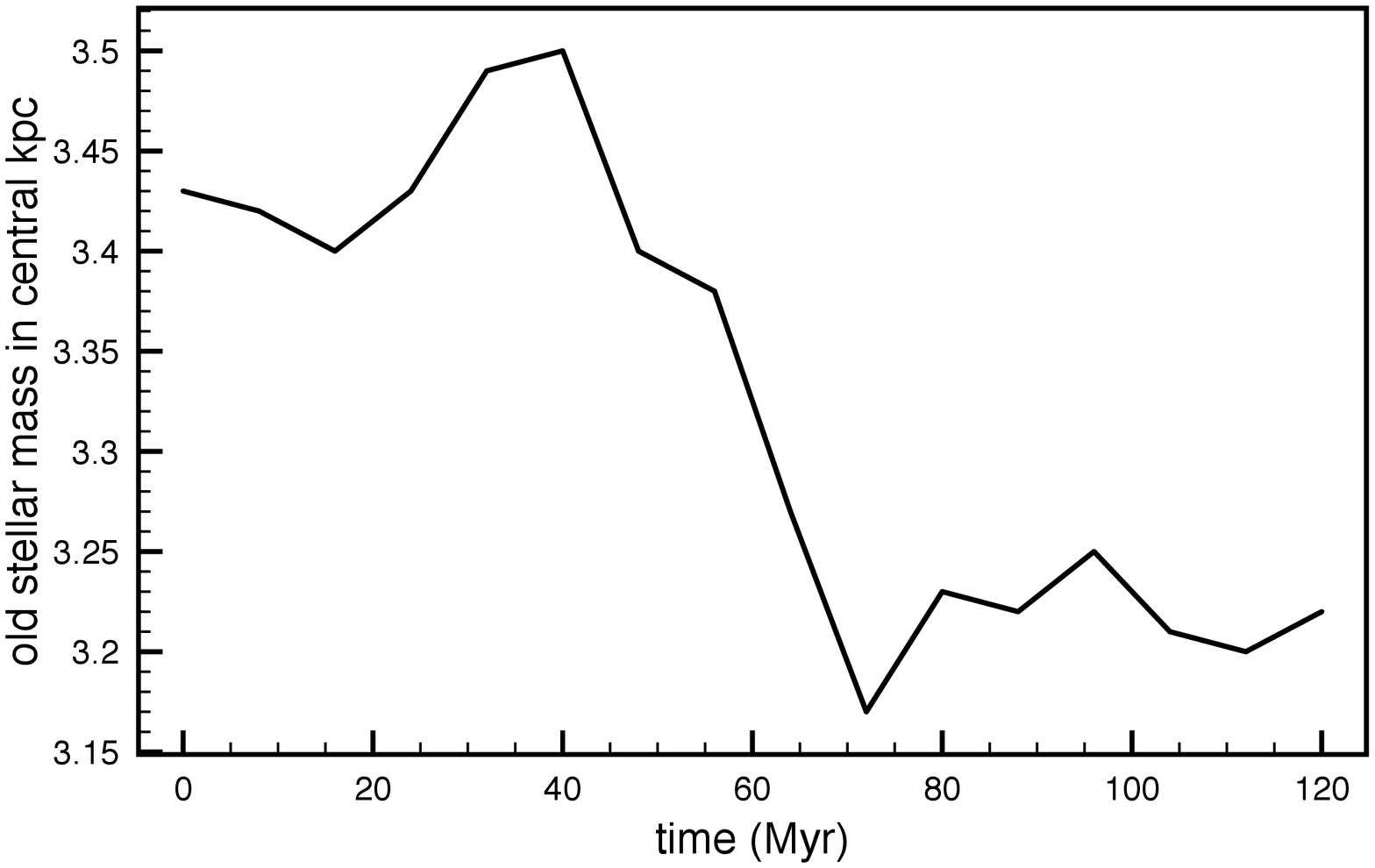}
\caption{Bulge evolution during the central coalescence of a giant clump (simulations from Bournaud et al., 2014). The panels display mass of gas in the central kpc (radius 500\,pc, top), the mass of stars younger than 100\,Myr plus the current time (i.e. younger than 100\,Myr at $t=0$ and younger than 150\,Myr at $t$=50\,Myr, middle), and the mass of stars older than 200\,Myr plus the current age (bottom, still in the central kpc). A massive clump coalesces with the bulge at $t\,\simeq\,60$\,Myr. The gas mass increases when the clump comes in, and decreases due to feedback-driven outflows. The mass of young stars in the bulge increases, but that of old star decreases, and the bulge mass is regulated to an almost constant value (even slightly decreasing in this example, although its S\'ersic index increases -- see text for details). This bulge regulation mechanism was proposed by Perret and collaborators (Perret et al., 2013, and Perret, PhD thesis, 2013).}
\label{fig_fb7}     
\end{figure}

This process of bulge self-regulation by the inflow of giant clumps is illustrated in a simulation similar to those in Bournaud et al. (2014) in Figure~\ref{fig_fb6}. In a galaxy of stellar mass $5.3\times 10^{10}$\,M$_\odot$ with 43\% of gas during the analysed period, a giant clump of $3.6\times10^{8}$\,M$_\odot$ of gas and $2.3 \times 10^{8}$\,M$_\odot$ of stars coalesces with the central bulge, the stellar mass of which is $3.3\times10^{9}$\,M$_\odot$. The clump brings gas to the central kpc, and young stars (from the clump stellar content and from central star formation in the clump gas). The local starburst consumes about 60\% of the clump gas within 25\,Myr, with a star formation rate in the central kpc peaking\footnote{After applying Gaussian time smoothing of FWHM 2\,Myr to erase fluctuations related to the numerical sampling of star formation.} at 13\,Myr in Figure~\ref{fig_fb6}. Note that this is a high surface density of star formation rate in the central kpc, but it does not drive the entire host galaxy outside of the typical scatter of the Main Sequence \citep{schreiber} and the entire host galaxy does not turn into a starburst. The gas mass in the central kpc then decreases by a larger amount than just gas consumption by star formation, under the effect of stellar feedback-driven outflows. The response to the sudden mass increase (clump accretion) and decrease (gas outflows) affects the stellar content of the bulge, with some aged stars leaving the bulge region toward an extended stellar halo on high-eccentricity orbits. In the particular example shown in Figure~\ref{fig_fb7}, the bulge gains about $2\times10^{8}$\,M$_\odot$ from the clump coalescence process, but rapidly looses $2.3\times10^{8}$\,M$_\odot$ of aged stars, thus slightly reducing the total bulge mass after this clump coalescence event. Detailed statistical studies remain to be performed, but the simulations studied in \citet{perret14} show that the bulge mass can be regulated to reasonable amounts that do not exceed 10-15\% of the stellar mass for Milky Way-mass galaxies, owing to the three regulation processes listed above -- 1: regulation of the clump mass by steady outflows and dynamical loss of aged stars, 2: regulation of the gas richness of the clumps by re-accretion of gas from the disk, and 3: regulation of the central bulge mass by central starbursts and relaxation during clump coalescence. A detailed accounting in cosmological context however remains required to study the bulge mass budget over the long-lasting clumpy unstable state from $z>3$ to $z \simeq 1$.


An alternative solution was recently proposed by \citet{combes-bulges} who demonstrated than in MOND dynamics, realistic clumpy disk are still predicted by simulations of $z \,=\,2$ galaxies, but the efficiency of bulge formation is lowered as the clump migration timescale increases. Gravitational torquing and inflows should still be present, but without central relaxation through clump coalescence, they may form only a pseudo-bulge.

\section{The associated growth of supermassive black holes}

\begin{figure}[!ht]
\centering
\includegraphics[width=11.5cm]{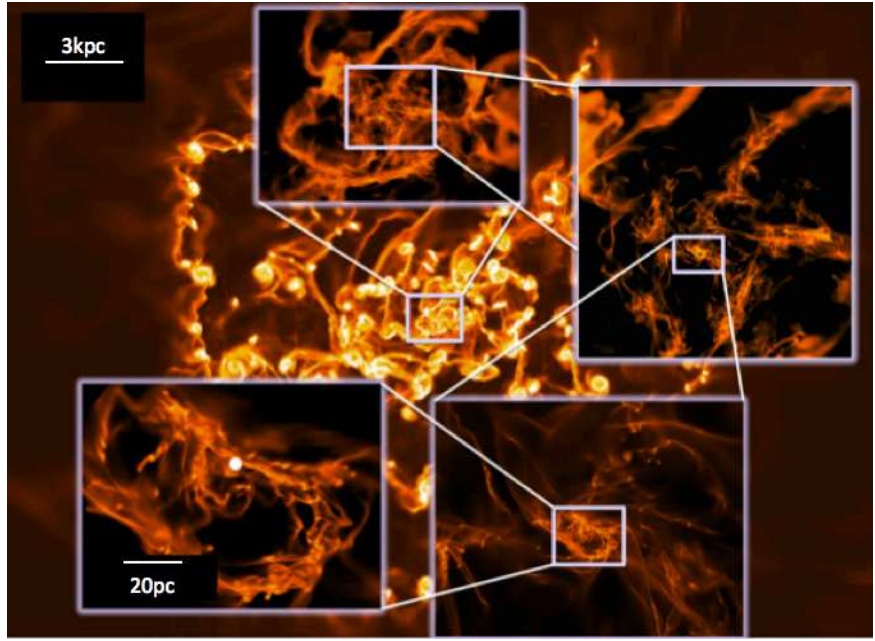}
\caption{AMR simulations of a high-redshift gas-rich disk galaxy with a central SMBH, using gradual zooms in the AMR refinement toward the central black hole, with a spatial resolution reaching 0.02\,pc in the innermost regions. The white circle in the last zoom represents the SMBH position and its Bondi radius. The disk instability drives steady gas inflows toward the SMBH. AGN feedback triggers hot winds that escape through low-density holes and leave the accreting channels almost unaffected. The large-scale star-formation activity also remains unaffected in spite of the efficient AGN-driven outflows. Figure courtesy of Jared Gabor.}
\label{fig_fb8}     
\end{figure}

The violent instability of high-redshift galaxies brings large amounts of gas toward their central regions. This is achieved through the migration of gas-rich clumps, and more generally by a global inflow of gas driven by gravity torques between the dense features arising from the instability, and compensating for the turbulent losses. Simulations of this process (Bournaud et al., 2011) have shown that an inflow of about 1 solar mass per year persists down to the central few parsecs, as the gas is not entirely depleted into star formation. It is then sufficient to have one percent of the mass brought to the central pc accreted by the central supermassive black hole (SMBH) to grow this SMBH in realistic proportions compared to the usual scaling relations for bulges and SMBHs. Various small-scale mechanisms in the central parsec can indeed lead 1\% of the available inflowing gas mass to be accreted by the SMBH \citep{review-bar-agn-correl}. The process was studied in detail in \citet{gabor-bh1} who has shown that bright Eddington-limited episodes of Active Galactic Nuclei (AGN) accretion can be triggered by the disk instability, and could contribute to the bulk of the supermassive black hole mass growth at $z=1-3$ for galaxies of stellar mass $10^{10-11}$\,M$_\odot$. The process is illustrated in Figure~\ref{fig_fb8}. In the broader cosmological context, the role of cold gas accretion onto gas-rich galaxies and internal instabilities was probed by \citet{dubois12,dubois13}.

Observationally, there is a general lack of correlation between the occurrence of AGN and morphological signatures of major mergers (e.g., \citealt{kocevski}) except in the most luminous QSOs found preferentially in major mergers. In fact, moderately bright AGN that drive the bulk of SMBH growth are mostly located in normally star-forming, Main Sequence galaxies \citep{mullaney}, which are generally clumpy unstable disks at $z=1-3$. Searches for a direct link between disk instabilities and AGN are hampered by the high gas column densities, typically a few times $10^9$\,M$_\odot$\,kpc$^{-2}$ in the central regions of these galaxies, sufficient to reach Compton thickness or at least severely attenuate the X-ray signatures of potential AGN, independently of their feeding mechanisms. Using optical line emission to probe AGN, \citet{bournaud12} have shown that at intermediate redshift ($z\approx 0.7$) there are more AGN in clumpy unstable disks, than in regular smooth disks of the same mass and size (both clumpy unstable disks and modern spiral types co-exist at such intermediate redshifts). At high redshift $z \, \geq \, 1$, \citet{trump} find an AGN frequency as high in clumpy disks as in compact early-type galaxies, which are known to be frequent AGN hosts (compared to star-forming spirals at low redshift), confirming the efficient feeding of AGN in clumpy disks. Yet a direct comparison of AGN feeding in clumpy unstable disks and in more ``stable'' disks is impossible at $z\,>\,1$, as stable spiral disks are virtually inexistent at $z \,>\,1$. 

\medskip

High-redshift disk instability can contribute to SMBHs in two other ways. First, the clumpy accretion onto black holes can help increase their mass more rapidly at early epochs, which subsequently increases the limiting Eddington rate and makes possible for the SMBH to grow its mass more rapidly. This potential solution to the problem of very massive and bright AGN at very high redshifts ($z \sim 6$, \citealt{tdm}) was studied in \citet{degraf}. Second, the clumps could be the formation site of SMBH seeds, if they form intermediate mass black holes through runaway stellar collisions, which their estimated star formation rate densities make possible. These seeds could be gathered centrally along with clump migration into an SMBH \citep{EBE08BH}, which was potentially supported by some observed spectral signatures \citep{shapiro09}.

 Studies of the response to feedback show that the feeding of AGN by disk instability does not quench star formation, and not even the fuelling of the AGN itself. AGN from clumpy disks produce high-velocity outflows that are collimated by the density and pressure gradients in the disk, and escape perpendicularly from the disk plane from the nuclear region \citep{gabor-bh2}. This is agreement with recent observations of high-velocity winds emerging preferentially from the nuclear regions \citep{FS14}. Outflows from star formation are more widespread above the entire disk and its star-forming clumps (observations: \citealt{newman}, simulations: \citealt{B14}, Hopkins et al., 2013). Hence the AGN feedback does not affect the inflowing, which is the fuel for future AGN feeding, and the extended gas disks including its star-forming regions. This holds even once long-range radiative effects are taken into account \citep{roos}. The AGN luminosity and accretion rate strongly fluctuate over Myr-long timescales, which results from the high heterogeneous, turbulent nature of the inflowing gas, rather than from the regulation by feedback \citep{gabor-bh2, degraf}.

\section{Disk instabilities and early-type galaxy formation}

The instability-driven inflow scales like the circular velocity squared (Section~3.4) and is thus much more intense in high-mass galaxies. Clump migration is also faster, following the dynamical friction timescale in massive galaxies with high-density disks and halos. This raises the question of whether the violent instability of high-redshift galaxies can form early-type galaxies (ETGs) at high masses, i.e. entirely spheroid-dominated systems rather than just bulges in the center of disk-dominated systems.

Theoretically, these strong inflows can lead to disk contraction in a timescale not larger than 1\,Gyr, and the instability-driven bulge growth rate could lead to a bulge-dominated system, through which the disk is stabilised and star formation is quenched \citep{DB14}, with the help of stellar spheroids stabilising gas disks to quench star formation \citep{martig09}. Recent cosmological simulations have probed these possible mechanisms, where the strong inflow first forms wet compact star-forming systems, which are subsequently quenched and turned into red compact ETGs \citep{ceverino14, zolotov}. The high S\'ersic indices and dispersion-dominated kinematics are consistent with these being the progenitors of modern ETGs. The transitions from the compact star-forming system to a quenched one could correspond to observations of the so-called ``blue nuggets'' and ``red nuggets'' at high redshift \citep{barro14}.

Observations of giant clumps or clump remnants in the innermost regions of young ETGs in the Hubble Ultra Deep Field \citep{E04etg} support this scenario. \cite{B11b} have also shown that mergers of gas-rich unstable disks lead to compact spheroid formation when the instability in the cold interstellar phase is taken into account during the merger. Nevertheless, it remains unknown whether these processes can explain the detailed phase space structure of modern ETGs, including the observed families of fast and slow rotators \citep{ems07} which could also be relatively well explained in the cosmological context without invoking a major role of disk instabilities \citep{naab}.

\section{Comparison to secular disk instabilities at lower redshift}

\begin{figure}[!ht]
\centering
\includegraphics[width=11.5cm]{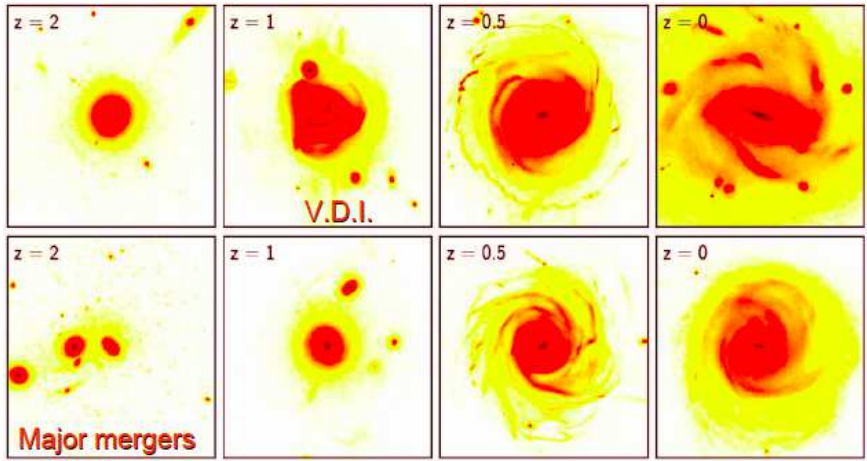}
\caption{Simulations in cosmological context, zoomed on individual galaxies (from Martig et al., 2012), displaying the stellar mass surface density (panel size: 20$\times$20\,kpc). These simulations show the transition from a ``violent phase'' at $z>1$ with violent disk instabilities (V.D.I., top) and giant clumps and, more rarely, merger-driven starbursts, to a ``secular phase'' with bars and spiral arms at $z\,<\,1$. Interestingly, this transition shows a ``downsizing'' behaviour with stellar mass, i.e. it occurs later-on for lower-mass galaxies, which could explain that clumpy disk instabilities persist longer for lower-mass galaxies (Elmegreen et al. 2007, Bournaud et al. 2012) and regular barred spiral morphologies arise earlier-on for high-mass galaxies (Sheth et al. 2008, Kraljic et al. 2012).}
\label{fig_fb9}     
\end{figure}

The violent instability where the entire disk is self-regulated at $Q \simeq 1$ persists until about redshift 1, for the galaxy masses that we have studied here \citep{elmegreen07, genel2, DSC09, ceverino14a}. This violent phase, with irregular clumpy disks, growing spheroids, and relatively frequent mergers, has a morphology poorly correlated to the final bulge/disk ratio of today's descendent galaxies \citep{martig12}. After $z \sim 1$, galaxies enter their secular phase where a stable thin disk grows and slowly evolves, with a bulge/disk ratio close to the final value. This regime differs by having globally $Q>1$, with $Q \leq 1$ only locally, for instance for gas compressed in spiral arms and in which small molecular clouds form by various local instabilities \citep{renaud13}. The evolution from the early violently unstable phase to the secular stable spiral disks is shown for two typical cases of zoom-in simulations in cosmological context in Figure~\ref{fig_fb9}.

The mild instability of modern spirals differ from the global instability of their high-redshift progenitors in various ways. Present-day disks globally develop only non-axisymmetric ($m\geq1$) modes such as spiral arms and bars, and gravitational collapse at $Q \leq 1$ can occur only locally in small over-densities of gas. The associated inflows are much slower, with only a few percent of the angular momentum transferred outwards per rotation period even in strongly barred galaxies \citep{combes-gerin,BCS05}. Mass inflows and vertical resonances can secularly grow central spheroids. Yet, simulations in cosmological context in \citet{kraljic12} suggest that the contribution of these low-redshift secular instabilities in the bulge mass budget is also more modest than for high-redshift instabilities, although the secular phase last longer and sometimes grows massive peanut-shaped bulges. The absence of violent relaxation, unlike the central coalescence of giant clumps, is such that the process mostly results in pseudo-bulges rather than classical bulges with high S\'ersic indices (Chapter~6.3). 

The inflow toward AGN and SMBH is also much more modest, but another difference here is the common presence of Inner Lindblad Resonances (ILRs) in low-redshift spiral galaxies -- high-redshift disks generally have no ILR associated to the clump instability. As a consequence, the inflow stops and the gas is stored at the ILR radius until a nuclear instability occurs and brings the gas reservoir inwards. The process can be repeated with cyclic AGN feeding, but a large fraction of the inflowing gas can also be depleted through star formation in the meanwhile \citep{ems15}. Indeed, observations point out that the correlation between galactic bars and AGN is complicated by a number of factors \citep[see e.g.][and references therein]{CoeGad11}. However, it can be argued that the correlation between nuclear bars and AGN is more straightforward \citep{review-bar-agn-correl}.

\section{Summary}

High-redshift star-forming galaxies at $z \, \simeq \, 1-3$ have irregular optical morphologies dominated by a few bright giant clumps, also faintly detectable in the near-infrared. There is broad evidence that these giant clumps (of a few 10$^{8-9}$\,M$_\odot$ of gas and stars and 500-1000\,pc diameter for the biggest ones) form mostly by in-situ gravitational instability in gas-rich, turbulent galactic disks. This is largely supported by photometry, kinematics, and stellar population studies. Recently, a first example of direct gravitational collapse of a giant clump could be directly probed \citep[in the form of a very massive collapsing star-forming blob with only very limited and very young stellar counterparts,][]{zanella}. Only a small fraction of clumps exhibit older stellar populations and may form ex-situ, in the form of small satellites of gaseous clumps that merge with the disk from the outside.

The modern understanding of galaxy formation in the standard cosmological framework explains the high gas fractions and resulting disk instability as the outcome of steady accretion of cosmological gas reservoirs (and some companion galaxies) at high mass rates, keeping the gas fraction high, the Toomre stability parameter low, and the disk in a globally unstable state. The disk increases its gas velocity dispersion (or turbulent speed) to self-regulated its dynamics in a steady state about $Q \simeq 1$.  

Hence the high-redshift progenitors of Milky Way-like spirals differ from modern disk galaxies, which have only weak non-axisymmetric instabilities (bars and spiral arms) and in which gas undergoes gravitational collapse only in limited regions, in the form of transient low-mass molecular clouds.


The giant clumps and the underlying instability can build a galactic bulge in several ways. The first one is the instability-driven inflows, which pumps gravitational energy into the interstellar turbulence cascade to compensate for the radiative losses. This inflow builds a central mass concentration in the form of a pseudo-bulge -- unless another process increases the relaxation and turns this central concentration into a classical bulge.

Detailed numerical models of star formation and feedback in a multi-phase ISM, and observations of stellar population ages, mostly support that the giant clumps can survive against feedback from young massive stars for a few hundreds of Myr, unlike nearby molecular clouds. In this case the giant clumps undergo dynamical friction from the host galaxy and its dark matter halo and migrate inward in a few $10^8$\,yr, coalesce with the central bulge, or form a bulge if no bulge is present yet. In this case the induced relaxation is generally found to turn the central spheroid into a classical bulge with a high S\'ersic index.

The detailed properties of bulges built by disk instabilities remain uncertain, and highly dependent on the physics of stellar feedback. Some studies find that it could be possible to form only a pseudo-bulge, but others find that in the most massive galaxies the whole system may turn into a classical spheroid through the violent disk instability, consistent with properties of early-type galaxies. 

An interesting property of clump migration and central coalescence is that the induced relaxation can affect the stellar orbits of a pre-existing bulge, and cause dynamical evaporation from the central bulge toward a very extended, low-density faint stellar halo. In this case the process gets self-regulated and the bulge mass is not grown beyond 10--15\% of the total stellar mass for a Milky Way-mass high-redshift galaxy.


The violent instability of high-redshift disk galaxies presents other interesting properties that can form other sub-galactic components concurrently with bulges. The instability-driven inflow can typically provide one solar mass per year toward the central parsec, which may be sufficient to dominate the feeding of central supermassive black holes in moderate mass galaxies, and this is potentially supported by observations of active galactic nuclei in Main Sequence galaxies. Along with bulges and central black holes, the violent instability of high redshift galaxies can also grow the old thick stellar disks, which are ubiquitous around present-day spiral galaxies.

\begin{acknowledgement}
I am grateful to Anna Cibinel, Jared Gabor, Marie Martig, Valentin Perret and Florent Renaud for providing some of the figures and material used in this review, to Avishai Dekel and Bruce Elmegreen for triggering many new studies on the instability of high-redshift galaxies and the associated growth of bulges, to the anonymous referee, and to Dimitri Gadotti for  detailed editing of the manuscript. The simulations shown in Figure~7 were carried out on the GENCI computing resources and TGCC/Curie, under allocation GENCI-2015-04-2192.
\end{acknowledgement}

\newcommand\mnras{MNRAS}
\newcommand\apj{ApJ}
\newcommand\apjs{ApJ Supp.}
\newcommand\apjl{ApJL}
\newcommand\aj{AJ}
\newcommand\nat{Nature}
\newcommand\aap{A\&A}

\end{document}